\documentclass[aps,pra,twocolumn,superscriptaddress,showpacs]{revtex4-1}
\usepackage{amsmath,amssymb,mathrsfs}
\usepackage{graphicx}
\usepackage{hyperref}
\usepackage[dvips]{color}

\def\avg#1{\langle#1\rangle}    \def\<{\langle}         \def\>{\rangle}

\def\up{\uparrow}       \def\down{\downarrow}

  \def\Bq{{\mathbf q}}
  \def\B0{{\mathbf 0}}

\def\be{\begin{equation}}       \def\ee{\end{equation}}
\def\bea{\begin{eqnarray}}      \def\eea{\end{eqnarray}}
\def\nn{\nonumber}

\begin{document}
\title{Finite temperature damping of collective modes of a BCS-BEC
  crossover superfluid}
\author{Zixu Zhang}
  \affiliation{Department of Physics and Astronomy, University of
  Pittsburgh, Pittsburgh, Pennsylvania 15260, USA}

\author{W. Vincent Liu}
  \affiliation{Department of Physics and Astronomy, University of
  Pittsburgh, Pittsburgh, Pennsylvania 15260, USA}
\affiliation{Center for Cold Atom Physics, Chinese Academy of
Sciences, Wuhan 430071, China}
\date{\today}

\begin{abstract}
A new mechanism is proposed to explain the puzzling damping of collective excitations, which was recently observed in the experiments of strongly interacting Fermi gases below the superfluid critical  temperature on the fermionic (BCS) side of Feshbach resonance.
Sound velocity, superfluid density and damping rate are calculated with effective field theory. We
find that a dominant damping process is due to the interaction between
superfluid phonons and thermally excited fermionic quasi-particles, in
contrast to the  previously proposed pair-breaking mechanism.
Results from our effective model are compared quantitatively with
recent experimental findings, showing a good agreement.
\end{abstract}


\maketitle

\section{Introduction}
In recent years, experimental studies on ultracold Fermi gases in
Bose-Einstein condensation (BEC) to Bardeen-Cooper-Schrieffer (BCS)
crossover \cite{NSR, PhysRevLett.71.3202, giorgini:1215} have shown many interesting results \cite{OHara12132002, PhysRevLett.92.040403, PhysRevLett.92.120403,  PhysRevLett.103.170402}. By varying external
magnetic field to effectively tune the scattering length through
Feshbach Resonance, or changing the temperature, collective behaviors,
such as oscillation frequency, sound velocity and damping rate, have
been extensively investigated
\cite{Bartenstein2004, PhysRevLett.94.170404, Joseph2007a,Wright2007a, Altmeyer2007}.
Recently, several experiments show a puzzling damping of collective
excitations  occurs in
superfluid regime
where the system is at a finite temperature and on the (fermion) BCS
side of the crossover
\cite{Bartenstein2004,Wright2007a,Altmeyer2007}.

In the theoretical front, the properties of collective
excitations in the crossover have
been investigated by a variety of established methods \cite{giorgini:1215},
including work that used the hydrodynamic theory \cite{0295-5075-65-6-749, PhysRevLett.93.040402, PhysRevLett.93.190403}, the Gaussian fluctuation approximation
from a microscopic Lagrangian \cite{PhysRevB.55.15153, PhysRevA.74.063626, PhysRevA.75.033609, PhysRevA.77.023626}, and
fermion-boson coupled model \cite{PhysRevLett.87.120406, PhysRevLett.89.130402, PhysRevA.67.063612}. The Boltzmann equation \cite{Vichi2000, PhysRevA.71.033607, PhysRevA.76.045602, PhysRevA.78.053609}  and semiclassical dynamics \cite{PhysRevA.73.013621, PhysRevA.78.053619} were also used to explain the frequency and damping rate of oscillation modes.

Effective field theory has also been used in some previous work to
study ultracold systems at zero temperature, including expanding the
goldstone fields to higher orders, or low energy expansion
\cite{PhysRevLett.96.080401, Valle20091136,Schakel2011193}. In this paper,
we use effective field theory to study the collective modes of the
superfluid state of an unpolarized strongly interacting cold Fermi gas
at finite temperatures near the Feshbach resonance regime. Our
effective theory is constructed based on a single-channel resonance
model, which is known to be adequate for a broad Feshbach
resonance. Furthermore, we focus on the BCS side of the crossover,
where puzzling damping phenomena have been observed and received much
attention~\cite{Bartenstein2004,Wright2007a,Altmeyer2007}. In this regime, the
BEC molecular effect is negligible, and hence the single-channel
calculation which we adopt is simple and valid. Our main results are
as follows. First, our theory shows that a dominant damping process is
due to interaction between superfluid phonons and thermally excited
fermionic quasi-particles at finite temperatures, in contrast to the
previously proposed pair-breaking mechanism. Such a process is related
to the well-known Landau damping, which was previously discussed in a semiclassical approach by taking the $\hbar \rightarrow 0$ limit of the time-dependent Hartree-Fock-Bogoliubov equations
\cite{PhysRevA.73.013621, PhysRevA.78.053619}.  Second, we calculate the damping
rate of collective excitations for the physical systems and find a
good comparison with the recent experimental findings. Throughout the
paper, the Planck constant $\hbar$ and Boltzmann constant $k_B$ are
set unity ($\hbar\equiv k_B\equiv 1$ in units).

\section{MEAN FIELD THEORY}
We consider an unpolarized ultracold fermionic system with two spin
species (spin $\uparrow$ and $\downarrow$) of equal mass. The
partition function of the system in the grand canonical ensemble can
be described in the path integral formalism
\begin{equation}
Z=\int D (\psi_\sigma^*, \psi_\sigma) \exp(-S_{\psi})
\label{eq:1st}
\end{equation}
where $S_\psi= \int dx \mathcal{L}_{\psi}(x)$ is the
action. $\psi_\sigma$ is the fermionic field for spin
$\sigma=\uparrow, \downarrow$. $x$ is a four vector
$x=(\mathbf{x},\tau)$, in which $\mathbf{x}$ is the spatial coordinate
and $\tau$ is imaginary time in the range $0<\tau<\beta$, $\beta= 1/T$
and $T$ is the temperature of the system.
The following Lagrangian is used to describe the system:
\begin{equation}
\mathcal{L}_{\psi}={\psi^*_\sigma \left(\partial _\tau -\frac{\nabla^2}{2m_a}-\mu \right) \psi_\sigma}+g\psi^*_\uparrow\psi^*_\downarrow\psi_\downarrow\psi_\uparrow,
\label{eq:1stlag}
\end{equation}
where $m_a$ is the mass of the fermion, $g$ is the interaction
(negative on the BCS side), and $\mu$ is the chemical potential. The
summation over $\sigma$ is implicit.

As usual, we use a Hubbard-Stratonovich transformation by introducing
an auxiliary complex bosonic field $\Delta(x)$ to eliminate the
quartic term in Eq.~\eqref{eq:1stlag} and get
\begin{equation}
Z=\int D (\psi_\sigma^*, \psi_\sigma) D(\Delta^*, \Delta) \exp(-S_{\psi,\Delta}),
\label{eq:hstransform}
\end{equation}
where the lagrangian is now
\begin{equation}
\mathcal{L}_{\psi,\Delta}=\psi^*_\sigma \left(\partial _\tau -\frac{\nabla^2}{2m_a}-\mu \right)\psi_\sigma+(\psi^*_\uparrow\psi^*_\downarrow\Delta+c.c) -\frac{|\Delta|^2}{g}.
\label{eq:mflag}
\end{equation}
In this transformation, the (auxiliary) field $\Delta(x)$ acquires
exactly the same expectation value as the pair field at the saddle
point:
$$
\avg{\Delta} = \avg{\psi_\down\psi_\up}\,.
$$
The fermionic field is now in quadratic form and can be integrated out to get an effective action for the field $\Delta$
\begin{equation}
Z=\int D(\Delta^*,\Delta) \exp (-S_\Delta),
\label{eq:mfeffective}
\end{equation}
where
\begin{equation}
S_{\Delta}=-\int d^4 x \left(\frac{|\Delta|^2}{g}+ \textrm{tr} \ln G^{-1} \right)
\label{eq:mfaction}
\end{equation}
and
\begin{eqnarray}
G^{-1}&=&
\left(
\begin{array}{cc}
\partial _{\tau}-\frac{\nabla ^2}{2m_a}-\mu & \Delta\\
\Delta^* & \partial _{\tau}+\frac{\nabla ^2}{2m_a}+\mu
\end{array}
\right).
\label{eq:mfg0}
\end{eqnarray}

In the mean field approximation, we seek for a spatially uniform saddle point solution $\Delta_0$ of $S_{\Delta}$, together with the requirement of fixing the number density $n$. These two conditions determine the mean field equations in the crossover
\begin{eqnarray}
\frac{\delta S_{\Delta}}{\delta \Delta_0}=0,\nonumber\\
\frac{\partial \ln Z}{\partial \mu}=n.
\label{eq:mfeqns}
\end{eqnarray}
In momentum space we get
\begin{eqnarray}
\frac{1}{g}+\sum_{\mathbf{k}}\frac{1}{2E_k} \tanh \left( \frac{\beta E_k}{2} \right)&=&0,\nonumber\\
\sum_\mathbf{k} \left[1-\frac{\xi_k}{E_k}\tanh \left(\frac{\beta E_k}{2} \right) \right]&=&n,
\label{eq:mfeqns2}
\end{eqnarray}
where $E_k=\sqrt{\xi_k^2+\Delta^2}$ and $\xi_k=\frac{k^2}{2m_a}-\mu$.

We use the usual regularization procedure,
\begin{equation}
\frac{m_a}{4\pi a}=\frac{1}{g(\Lambda)}+\sum_{|\mathbf{k}|<\Lambda}
\frac{1}{2\epsilon_k}\,,
\label{eq:mfreg}
\end{equation}
where $\epsilon_k=|\mathbf{k}|^2/2m_a$ and $a$ is the effective
scattering length.  Taking $\Lambda\rightarrow\infty$,
Eq.~\eqref{eq:mfeqns2} can be expressed in terms of $a$ as
\begin{eqnarray}
\frac{m_a}{4\pi a}&=&\sum_{\mathbf{k}} \left[\frac{1}{2\epsilon_k}-\frac{1}{2E_k} \tanh \left( \frac{\beta E_k}{2} \right) \right],\nonumber\\
n&=&\sum_\mathbf{k} \left[1-\frac{\xi_k}{E_k}\tanh \left(\frac{\beta E_k}{2} \right) \right].
\label{eq:mfreg2}
\end{eqnarray}

By self consistently solving Eq.~\eqref{eq:mfreg2}, we can get
$\Delta_0$ and $\mu$ as function of $1/k_F a$ and $T/T_F$ in mean
field level, where $k_F$ and $T_F$ are the Fermi momentum and
temperature of the non-interacting Fermi gas in free space with the
same number density $n$. In the following sections, we will assume the
state of the system is already superfluid and will use the mean field
results, such as the gap parameter, as input to calculate the physical
quantities of excitations.

\section{Quantum Fluctuations}
To consider the quantum fluctuations, we no longer treat the order parameter $\Delta$ as a homogeneous constant in Eq.~\eqref{eq:mflag}, but write it as
\begin{equation}
\Delta(x)=\Delta_0\left(1+\lambda(x)\right) e^{i2\varphi (x)}.
\label{eq:qfdecompose}
\end{equation}
$\lambda$ is the amplitude fluctuation and $\varphi$ is the phase
fluctuation around the saddle point
solution $\Delta_0$, and both of them are real. Then one can apply a
local U(1) gauge transformation to a new gauge where the order
parameter is real everywhere in space
\begin{equation}
\Delta=\tilde{\Delta}e^{i2\varphi(x)}, \psi _\sigma(x)=\tilde{\psi}_\sigma(x)e^{i\varphi(x)},
\label{eq:phaseout}
\end{equation}
and $\tilde{\Delta}=\Delta_0 (1+\lambda)$ is real now~\footnote{We treat the Jacobian as a constant using the same approximation in Ref.~\cite{PhysRevA.77.023626, PhysRevB.62.6786}.}. In the new gauge, the Lagrangrian density Eq.~\eqref{eq:mflag} becomes
\begin{eqnarray}
\mathcal{L}_{\tilde{\psi},\varphi,\lambda} &=&\tilde{\psi}^*_\sigma(\partial_\tau- \frac{\nabla^2}{2m_a}-\mu)\tilde{\psi}_\sigma+ (\Delta_0\tilde{\psi}^*_\uparrow\tilde{\psi}_\downarrow^* +c.c.)\nonumber\\
&&+\tilde{\psi}^*_\sigma \tilde{\psi}_\sigma [i\partial_\tau\varphi+\frac{(\nabla\varphi)^2} {2m_a}]+\nabla\varphi\cdot\hat{\mathbf{J}}_\sigma\nonumber\\
&&+(\Delta_0\lambda\tilde{\psi}^*_\uparrow\tilde{\psi}_\downarrow^*+c.c.) -\frac{\Delta_0^2\lambda^2}{g}
\label{eq:1steffLag}
\end{eqnarray}
where $\hat{\mathbf{J}}_\sigma=\hat{\mathbf{J}}_\sigma(x)=-\frac{i}{2m_a} [\tilde{\psi}^*_\sigma(\nabla\tilde{\psi}_\sigma)-(\nabla\tilde{\psi}^*_\sigma) \tilde{\psi}_\sigma]$ is the fermion current field.

We now integrate out the fermionic field which is gapped. In momentum
space, for the fermionic field $\tilde{\psi}$, we use
$k=(\mathbf{k},i\omega_m)$, where $\omega_m=(2m+1)\pi/\beta$ is the
fermionic Matubara frequecy and $m$ is integer. For the bosonic field
$\lambda$ and $\varphi$, we use $q=(\mathbf{q}, i\omega_n)$, and
$\omega_n =2n\pi/ \beta$, where $n$ is integer.

Introducing two component spinor in momentum space
$\Psi_k^*=(\tilde{\psi}_{k\uparrow}^*,\tilde{\psi}_{-k\downarrow})$,
we can rewrite the action in momentum space as $S=\sum_{k_1,k_2}
\Psi_{k_1}^*G^{-1}(k_1,k_2)\Psi_{k_2}$, where
$G^{-1}=G_0^{-1}+\chi_1+\chi_2$ is $2\times2$ matrix. $G_0$, $\chi_1$,
and $\chi_2$ are given by
\begin{eqnarray}
G_0(k_1,k_2)&=&
\frac{\delta_{k_1,k_2}}{\omega_{m_1}^2+E_{k_1}^2}\left(
\begin{array}{cc}
i\omega_{m_1}+\xi_{k_1} & \Delta_0\\
\Delta_0 & i\omega_{m_1}-\xi_{k_1}
\end{array}
\right)\nonumber\\
\chi_2(k_1,k_2)&=&
-\sum_{q} \frac{\mathbf{q}\cdot(\mathbf{k_1-k_2-q})}{2m_a\beta V} \varphi_q \varphi_{k_1-k_2-q} \left(
\begin{array}{cc}
1 & 0 \\
0 & -1
\end{array}
\right)\nonumber\\
\chi_1(k_1,k_2)&=& \left(
\begin{array}{cc}
\chi_1^{11}(k_1,k_2) & \chi_1^{12}(k_1,k_2) \\
\chi_1^{21}(k_1,k_2) & \chi_1^{22}(k_1,k_2)
\end{array}
\right)
\label{eq:g0chi1chi2}
\end{eqnarray}
with
\begin{eqnarray}
\chi_1^{11}(k_1,k_2)&=&\left[ \frac{\omega_{m_1}-\omega_{m_2}}{\sqrt{\beta V}}+\frac{i(\mathbf{k_1^2-k_2^2})}{2m\sqrt{\beta V}} \right] \varphi_{k_1-k_2} \nonumber\\
\chi_1^{22}(k_1,k_2)&=&\left[ -\frac{\omega_{m_1}-\omega_{m_2}}{\sqrt{\beta V}}+\frac{i(\mathbf{k_1^2-k_2^2})}{2m\sqrt{\beta V}} \right] \varphi_{k_1-k_2}\nonumber\\
\chi_1^{12}(k_1,k_2)&=&\chi_1^{21}(k_1,k_2)=\frac{\Delta_0}{\sqrt{\beta V}}\lambda_{k_1-k_2}.
\label{eq:chi1}
\end{eqnarray}

Keeping the $\lambda$ and $\varphi$ field to quadratic order~\footnote{It means we adopt Gaussian approximation for the amplitude field $\lambda$ and low energy expansion $\varphi$ for the superfluid phase field.}
\begin{equation}
\int D(\Psi^*, \Psi)e^{-S_{\Psi, \lambda, \varphi}} \approx e^{\textrm{tr} [G_0 \chi_2-\frac{1}{2}G_0 \chi_1 G_0 \chi_1]}=e^{-S'},
\label{eq:momfinish}
\end{equation}
we get the effective action
\begin{eqnarray}
S'=S_{\varphi}+S_{\lambda}+S_{\varphi, \lambda},
\label{eq:2ndeffLag}
\end{eqnarray}
where
\begin{eqnarray}
S_{\varphi}&=&\sum_{q}\varphi_q\varphi_{-q}[f_1(q)\omega_n^2+[f_4(q)+\frac{n}{2 m_a}] \mathbf{q}^2],\nonumber\\
S_{\lambda}&=&\sum_{q} \lambda_q \lambda_{-q} f_3(q),\nonumber\\
S_{\varphi, \lambda}&=&\sum_{q} [f_2(q) \lambda_q \varphi_{-q} (-\omega_n-\frac{i}{m_a}\mathbf{q \cdot k})\nonumber\\
&&- f_2(q) \lambda_{-q}\varphi_q (-\omega_n+\frac{i}{m_a} \mathbf{q \cdot k})],
\label{eq:s1s2s3}
\end{eqnarray}
and~\footnote{To calculate $\textrm{tr} (G_0 \chi_2)$, convergence factors are needed. See Ref.~\cite{PhysRevA.77.023626}.}
\begin{eqnarray}
f_1&=&\frac{1}{\beta V} \sum_{k} \frac{\omega_m\omega_{m+n}-\xi_k\xi_{k+q}+\Delta_0^2} {(\omega_m^2+E_k^2)(\omega_{m+n}^2+E_{k+q}^2)},\nonumber\\
f_2&=&\frac{\Delta_0^2}{\beta V} \sum_{k} \frac{i\omega_m+i\omega_{m+n}+\xi_k+\xi_{k+q}}{(\omega_m^2+E_k^2) (\omega_{m+n}^2+E_{k+q}^2)}, \nonumber\\
f_3&=&-\frac{\Delta_0^2}{g}-\frac{\Delta_0^2}{\beta V} \sum_k \frac{\omega_m\omega_{m+n} +
\xi_k\xi_{k+q}-\Delta_0^2}{(\omega_m^2+E_k^2)(\omega_{m+n}^2+E_{k+q}^2)} ,\nonumber\\
f_4&=& \frac{1}{\beta V} \sum_{k} \frac{k^2 \cos ^2 \theta}{m_a^2} \frac{-\omega_m\omega_{m+n}+ \xi_k\xi_{k+q}+\Delta_0^2}{(\omega_m^2+E_k^2)(\omega_{m+n}^2+E_{k+q}^2)}. \nonumber\\
\label{eq:f1tof5}
\end{eqnarray}

The next step is to integrate out the amplitude fluctuation field
$\lambda$ and keep only the phase field. Define
\begin{eqnarray}
C_1(q)&=&f_1+f_2^2/f_3,\nonumber\\
C_2(q)&=&f_4+\frac{n}{2m_a},
\label{eq:c1c2}
\end{eqnarray}
and the effective action of phase field $\varphi$ is obtained as
\begin{equation}
S_{\varphi}=\sum_{q}\varphi_q\varphi_{-q} [C_1(q)\omega_n^2+C_2(q)\mathbf{q}^2].
\label{eq:c1c2expression}
\end{equation}

\section{Superfluid density and sound velocity}
Some physical quantities such as superfluid density and sound velocity can be evaluated by our theory. To get a full expression for the superfluid density, one can apply a Galilean boost on the order parameter in Eq.~\eqref{eq:phaseout}, and get the superfluid density by the shift of free energy in the superfluid hydrodynamic model with normal and superfluid components \cite{PhysRevA.74.063626, PhysRevA.75.033609, taylorthesis}. The total normal density includes bosonic fluctuation part and non-condensed fermionic Bogoliubov quasi-particle part. From the calculations of Ref.~\cite{PhysRevA.74.063626, PhysRevA.75.033609, taylorthesis}, the main contribution to the normal density on the BEC side is the bosonic fluctuation part, while on the BCS side, the main contribution is from the non-condensed fermionic quasi-particle part. Since our main focus is the damping on the BCS side, we will neglect the part of bosonic fluctuations in the normal  density in the following calculation.

To the zeroth order of $q$, or in the long wave length limit $q
\rightarrow 0$, all coefficients $f_1, f_2, f_3$, and $f_4$ are
real, in terms of which superfluid density and sound velocity can be
expressed. Since the imaginary part is zero, there is no damping for
the collective modes, to the lowest order.  $C_1' \equiv \lim_{q \rightarrow 0}C_1$ is related to the
density of states, and $C_2' \equiv \lim_{q \rightarrow 0}C_2
=n_s/2m_a$ is related to superfluid
density $n_s$~\cite{bensimons} \footnote{In $q \rightarrow 0$ limit, we automatically neglect the bosonic fluctuation part in the normal density}.

Take the limit of $q \rightarrow 0$ in the coefficients
Eq.~\eqref{eq:f1tof5}, and define $f_1'\equiv\lim_{q \rightarrow 0}
f_1$ and so on. After carrying out the Matsubara summation for
$i\omega_m$,  we get~\footnote{  We used the mean field gap equation
  Eq.~\eqref{eq:mfeqns} to get $f_3'$. Including corrections due to quantum fluctuations may change its form. For detailed discussions on the effect of quantum fluctuations, see Ref.~\cite{PhysRevA.77.023626, taylorthesis}.  }
\begin{eqnarray}
f_1'&=&\frac{1}{V}\sum_{\mathbf{k}}[\beta n_k(1-n_k)\frac{\xi
    _k^2}{E_k^2}+\frac{\Delta_0^2}{2E_k^3}(1-2n_k)]\,, \nonumber\\
f_2'&=&\frac{\Delta^2_0}{V}\sum_{\mathbf{k}}[\beta n_k(n_k-1)\frac{\xi_k}{E_k^2}+(1-2n_k)\frac{\xi_k}{2E_k^3}]\,,\nonumber\\
f_3'&=&\frac{\Delta_0^4}{V}\sum_{\mathbf{k}}[\beta n_k (n_k-1)\frac{1}{E_k^2}+\frac{1}{2E_K^3}\tan \frac{\beta E_k}{2}]\nonumber\,,\\
f_4'&=&\sum_{\mathbf{k}} \frac{\beta}{V}\frac{k^2
    \cos^2\theta}{m_a^2}n_k(n_k-1)\,, \nonumber\\
\label{eq:coeffreal}
\end{eqnarray}
where
$$n_k={1\over e^{\beta E_k} +1 } $$
is the quasi-particle Fermi-Dirac distribution. Eq.~\eqref{eq:c1c2expression} becomes
\begin{equation}
S=\sum_{q}\varphi_{q}\varphi_{-q}[C_1'\omega_n^2+C_2'\mathbf{q}^2]
\label{eq:effectivevarphi}
\end{equation}
with
\begin{eqnarray}
C_1'&=&f_1'+f_2'^2/f_3'\,,\nonumber\\
C_2'&=&f_4'+\frac{n}{2m_a}\,.
\label{eq:c1c2realpart}
\end{eqnarray}
After applying a Wick rotation $i\omega_n=\omega_q+i0^+$,  the zero of
the action,
Eq.~\eqref{eq:effectivevarphi}, gives the spectrum
$\omega_q=v_s|\mathbf{q}|$, where $v_s$ is the sound velocity.
Thus, we get the following physical quantities,
\begin{eqnarray}
\frac{n_s}{2m_a}&=&f_4'+\frac{n}{2m_a}\,,\nonumber\\
v_s^2&=&\frac{f_4'+n/2m_a}{f_1'+f_2'^2/f_3'}\,.
\label{eq:quatities}
\end{eqnarray}

At zero temperature, Eq.~\eqref{eq:coeffreal} reduces to
\begin{eqnarray}
f_{1,T=0}'&=&\frac{\Delta_0^2}{V} \sum_{\mathbf{k}}\frac{1}{2E_k^3},\nonumber\\
f_{2,T=0}'&=&\frac{\Delta^2_0}{V}\sum_{\mathbf{k}} \frac{\xi_k}{2E_k^3},\nonumber\\
f_{3,T=0}'&=&\frac{\Delta_0^4}{V}\sum_{\mathbf{k}}\frac{1}{2E_k^3},\nonumber\\
f_{4,T=0}'&=&0.\nonumber\\
\label{eq:zeroTcoeffreal}
\end{eqnarray}
Perform the summation for $f^\prime$'s over $\mathbf{k}$ as
integration in the thermodynamic limit,
and define the following integral quantities
\begin{eqnarray}
J_2&=&\int_0^{\infty} dk \frac{k^2}{E_k^3},\nonumber\\
J_4&=&\int_0^{\infty} dk \frac{k^4}{E_k^3},\nonumber\\
J_{\xi}&=&\int_0^{\infty} dk \frac{k^2\xi_k}{E_k^3}.
\label{eq:zeroTJs}
\end{eqnarray}
After some calculations, the sound velocity at $T=0$ is
\begin{equation}
v_{s,T=0}^2=\frac{1}{3m_a^2} \frac{J_2J_4\Delta_0^2}{J_2^2\Delta_0^2+J_{\xi}^2},
\label{eq:zeroTsound}
\end{equation}
which reproduces the result of Ref.~\cite{PhysRevA.74.042717}, albeit
that was derived in a different method.

\begin{figure}[t]
\includegraphics[width=240pt,height=97pt]{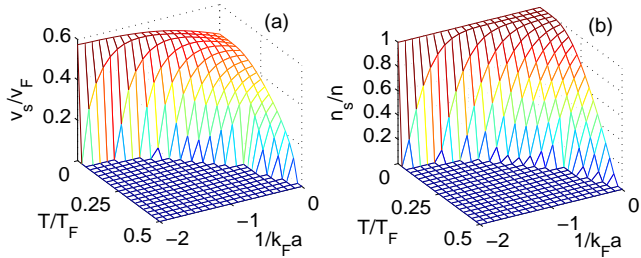}
\caption{(Color online) (a) Superfluid sound velocity compared with Fermi velocity $v_s/v_F$ in superfluid regime. (b) Superfluid density compared with the total density $n_s/n$.}
\label{fig:dandv}
\end{figure}
Fig.~\ref{fig:dandv} shows the numerical plot of superfluid density
and superfluid sound velocity with varying temperature and
scattering length in the BCS regime.

\section{Damping}
\subsection{Approximate formulas of the damping rate}

We start from the effective action, Eq.~\eqref{eq:c1c2expression}, for
the superfluid phase field. Keeping $q$ small but finite when
evaluating the coeffiencets $C_1$ and $C_2$ in Eq.~\eqref{eq:c1c2},
imaginary terms appear in Eq.~\eqref{eq:f1tof5}, which corresponds to
the damping of the collective modes. To get exact dispersion relation
including damping for the superfluid phonons, one needs to
self-consistently solve for poles of the effective action
Eq.~\eqref{eq:c1c2expression} of $\varphi$. However, in the regime where the
damping is small, we can simplify the calculation.

\begin{figure}[t]
\includegraphics[width=195pt,height=107pt]{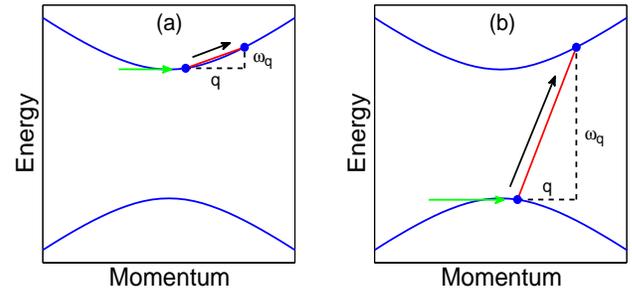}
\caption{(Color online) Two separate damping channels. The blue curves
are the fermionic quasi-particle spectrum. At $T=0$, the lower branch
is fully occupied and the higher branch is empty. The green arrow is
the incident phonon. The black arrow indicates the excitation of a
fermionic quasi-particle. The red straight line is the linear spectrum
of the phonon, where the slope is the sound velocity. The two black
dashed lines are the changes of momentum and energy of fermionic
quasi-particle, as indicated. (a) A phonon scatters an existing (by
finite temperature) Fermi quasi-particle to a different state in the
same energy branch. Such a process is known as Landau damping. This process can also happen in the
lower branch at finite temperature, which is not shown here. (b) A
phonon creates a quasi-particle and -hole pair across the lower and
upper bands, equivalent to Cooper pair breaking in the representation
of original particles.}
\label{fig:channel}
\end{figure}

The dispersion relationship is determined by setting the quadratic field term Eq.~\eqref{eq:c1c2expression} to be zero.
To evaluate the damping rate we apply a Wick rotation $i
\omega_n=\omega_0-i\gamma$ as follows,
\bea
&& -C_1(\tilde{q})(\omega_0-i\gamma)^2+
C_2(\tilde{q})\mathbf{q}^2 =0, \\
&& \tilde{q} \equiv (\Bq,\  i\omega_n\rightarrow \omega_0 +i0^+)\,, \nn
\eea
where $\omega_0$ is the oscillation
frequency, or the real part of the phonon mode, and $\gamma$ is the
damping rate. Although in principle, we should also keep $\gamma$ when
applying  the Wick
rotation to the $\omega_n$'s in coefficients $f_1, f_2, f_3$, and $f_4$, for small damping it is sufficient to just apply $i
\omega_n=\omega_0+i0^+$ for the coefficients to get the lowest order
results. This approximation is similar to that adopted in the
calculation for self-energy, for example,
Ref.~\cite{PhysRevLett.79.4056}.

For the numerators in Eq.~\eqref{eq:f1tof5}, we can still
set $q=0$. The reason to do this is that the numerators are all real,
and hence the small $q$ expansion gives only high order corrections to
the zeroth order contribution. If we  want only the leading order in
the long wavelength expansion of the damping rate, the high order
corrections in the numerators can be ignored, as we have checked.

Also, we can still use the results Eq.~\eqref{eq:coeffreal} obtained from last section for
the real parts of $C_1, C_2, f_1, f_2, f_3$, and $f_4$, because we are
not interested in the high order corrections to the real parts of
coefficients, which just give higher order correction to the damping
rate.

Adopting the above approximation, we can write $C_1=C_1'+iC_1'',
C_2=C_2'+iC_2'', f_1=f_1'+if_1''$ and so on, to find the poles of the
effective action
\begin{equation}
(C_1'+iC_1'')(\omega_0-i\gamma)^2-(C_2'+iC_2'')\mathbf{q}^2=0\,.
\label{eq:dampingc1c2a}
\end{equation}
Solving this equation gives the damping rate
\begin{equation}
\frac{\gamma}{\omega_0}=\frac{1}{2}\left(\frac{C_1''}{C_1'}- \frac{C_2''}{C_2'}\right)\,.
\label{eq:dampingc1c2b}
\end{equation}
From Eq.~\eqref{eq:c1c2}, assuming that all the imaginary parts are
small, we have
\begin{eqnarray}
C_1''&=&f_1''+2\frac{f_2'}{f_3'}f_2''-\left(\frac{f_2'}{f_3'}\right)^2 f_3''\,,\nonumber\\
C_2''&=&f_4''\,.
\label{eq:imc1c2}
\end{eqnarray}

After some calculation, we find that the imaginary parts of the
coefficients $f^{\prime\prime}$'s can be separated into
two channels $a$ and $b$, where in channel $a$
\begin{eqnarray}
f_{1,a}''&=&\frac{\pi}{V} \sum_{\mathbf{k}}(1-\frac{\Delta_0^2}{E_k^2})\delta[\omega_0-(E_{k+q}-E_k)] (n_k-n_{k+q})\,,\nonumber\\
f_{2,a}''&=&-\frac{\pi \Delta_0^2}{V} \sum_{\mathbf{k}}\frac{\xi_k}{E_k^2} \delta[\omega_0-(E_{k+q}-E_k)] (n_k-n_{k+q})\,,\nonumber\\
f_{3,a}''&=&-\frac{\pi \Delta_0^4}{V} \sum_{\mathbf{k}}\frac{1}{E_k^2} \delta[\omega_0-(E_{k+q}-E_k)] (n_k-n_{k+q})\,,\nonumber\\
f_{4,a}''&=&-\frac{\pi}{V} \sum_{\mathbf{k}}\frac{\mathbf{k}^2\cos^2 \theta}{m_a^2} \delta[\omega_0-(E_{k+q}-E_k)] (n_k-n_{k+q})\,,\nonumber\\
\label{eq:imf1tof5a}
\end{eqnarray}
and in channel $b$
\begin{eqnarray}
f_{1,b}''&=&\frac{\pi}{V}
\sum_{\mathbf{k}}\frac{\Delta_0^2}{2E_k^2}\delta(\omega_0-E_k-E_{k+q})
(1-n_k-n_{k+q})\,, \nonumber\\
f_{2,b}''&=&\frac{\pi \Delta_0^2}{V}
\sum_{\mathbf{k}}\frac{\xi_k}{2E_k^2} \delta(\omega_0-E_k-E_{k+q})
(1-n_k-n_{k+q})\,, \nonumber\\
f_{3,b}''&=&-\frac{\pi \Delta_0^2}{V}
\sum_{\mathbf{k}}(\frac{1}{2}-\frac{\Delta_0^2}{2E_k^2})
\delta(\omega_0-E_k-E_{k+q})\,, \nonumber\\
&& \times (1-n_k-n_{k+q})\,, \nonumber\\
f_{4,b}''&=&0\,.\nonumber\\
\label{eq:imf1tof5b}
\end{eqnarray}
For the expressions above, we would like to
remind the reader of the definitions of
$$E_k=\sqrt{\xi_k^2+\Delta^2} \quad \mbox{and} \quad
n_k={1\over e^{\beta E_k} +1 }.$$

The physical meaning of channel $a$ and $b$, as defined in the
above expressions, is illustrated in Fig.~\ref{fig:channel}.
Channel $a$ (Fig.~\ref{fig:channel}(a)) is the Landau damping. A (fermionic) quasi-particle absorbs or emits a
superfluid phonon and becomes another quasi-particle state at
a different momentum within the same energy band. In
channel $b$ (Fig.~\ref{fig:channel}(b)), a superfluid phonon excites a quasi-particle from the
lower band to the upper band,
creating a particle-hole pair excitation
in the quasi-particle eigenvector basis, which corresponds to a Cooper
pair breaking process in terms of original fermions.
Both channels cause the superfluid phonons (collective exicitations of
the superfluid state) to decay.

Before applying our theory to real experimental systems, let us discuss the properties of the damping in the $\omega_0 \rightarrow 0$ limit (with $\omega_0/|\mathbf{q}|=v_s$ fixed) at low temperature, where channel $b$ vanishes. To satisfy the $\delta$ function in Eq.~\eqref{eq:imf1tof5a}, we need to have
\begin{equation}
|\cos \theta|=\left|\frac{E_k m v_s } {\xi_k k} \right| \le 1,
\label{eq:analytical1}
\end{equation}
which is difficult to solve analytically.

We can still get some analytical properties if we combine the numerical and analytical analysis. At $1/k_F a=-0.45$ and $T/T_F<0.1$, numerical analysis shows that $k<{k_\mu}$ does not contribute to the integral. For $k>k_{\mu}$, we can substitute $k=k_{\mu}$ in Eq.~\eqref{eq:analytical1}, except for $\xi_k$, to get the range of $\xi_k$. We then apply a one-step iteration to get a more precise range for $\xi_k$. To satisfy Eq.~\eqref{eq:analytical1}, we need to have
\begin{eqnarray}
\xi_k&>&\frac{\sqrt{\xi_0^2+\Delta^2} m v_s}{\sqrt{2m(\xi_0+\mu)}} \equiv \xi_1,\nonumber\\
\xi_0&=&\frac{\Delta m v_s}{k_\mu}.
\label{eq:analytical2}
\end{eqnarray}

The next step is to write
\begin{eqnarray}
&&n_{k}-n_{k+q}\nonumber\\
&\approx& \frac{\partial n_k}{\partial E_k} \cdot (E_k-E_{k+q})\nonumber\\
&=&\beta \omega_0 n_k(1-n_k)\nonumber\\
&\approx& \beta \omega_0 e^{-\beta E_k}.
\label{eq:analytical3}
\end{eqnarray}
The second line comes from $\omega_0 \ll T$ and the fourth line comes from $T \ll \Delta$. Also, numerical analysis shows that the contribution from $C_1''$ is much smaller than that from $C_2''$ in the above regime. In addition, at low temperature, $C_2'\approx n/2 m_a$ since the normal density is negligible. Taking all the above into account, we get
\begin{equation}
\frac{\gamma}{\omega_0} \approx \frac{3\pi}{4}\frac{v_s^3}{v_f^3} \frac{1}{T} \int_{E_k>\sqrt{\xi_1^2+\Delta ^2}} dE_k \frac{E_k^4}{\xi_k^4} e^{-\beta E_k}.
\label{eq:analyticalbetter}
\end{equation}
The integral in Eq.~\eqref{eq:analyticalbetter} needs to be evaluated numerically in general cases. In the extremely low temperature, we can approximately treat $E_k^4/\xi_k^4$ in Eq.~\eqref{eq:analyticalbetter} as a constant and substitute in $\xi_k=\xi_1$, which gives
\begin{equation}
\frac{\gamma}{\omega_0} \approx \frac{3\pi}{4}\frac{v_s^3}{v_f^3} \frac{(\xi_1^2+\Delta^2)^2}{\xi_1^4} e^{-\frac{\sqrt{\xi_1^2+\Delta^2}}{T} }.
\label{eq:analyticalgood}
\end{equation}

An interesting feature of damping is that in Eq.~\eqref{eq:analyticalbetter} and~\eqref{eq:analyticalgood}, $\gamma / \omega_0$ is independent of $\omega_0$ when $\omega_0 \ll T$. Therefore, we should observe the damping even in the $\omega_0 \rightarrow 0$ limit in finite temperature. A physical understanding of this feature is that the damping from channel $a$ is the coupling between superfluid sound (phonon) and thermally excited Fermi quasi-particles, so that
\begin{eqnarray}
\gamma &\propto& \int dk f(k,\Delta, \mu) (n_k-n_{k+q})\nonumber\\
&\propto&  \omega_0 \int dk f(k,\Delta, \mu) \frac{d n_k}{d E_k},
\label{eq:gammaprop}
\end{eqnarray}
where $f(k,\Delta, \mu)$ is an arbitrary function independent of $\omega_0$. This provides an experimental method to verify our theory, which will be discussed later.

In the following section, we will apply our theory to real experimental systems. We will focus on the regime of $\gamma/ \omega_0 \ll 1$ on the BCS side of the Feshbach resonance, where our approximations are quantitatively controlled.

\subsection{Comparison with experiments}
The above theory can be applied to real physical
systems~\footnote{One should notice that in experiments, the cold
gases are trapped in anisotropic harmonic potentials. The parameters
used here, i.e., the Fermi energy, chemical potential, etc.,
correspond to the values at the center of the trap.}. We apply our
method to the experimental configuration in Ref.~\cite{Wright2007a}
and calculate the damping rate as function of both temperature and
scattering length. The number of particles is $N=4\times 10^5$. The
trapping frequencies are $\omega_x=2\pi\times830\ \textrm{Hz}$,
$\omega_y=2\pi\times415\ \textrm{Hz}$ and $\omega_z=2\pi\times22\
\textrm{Hz}$. Thus, $\bar {\omega}=(\omega_x \omega_y
\omega_z)^{1/3}=2 \pi\times 196\ \textrm{Hz}$. Reading from
Ref.~\cite{Wright2007a}, the oscillation (phonon) frequency is
$\omega_0 \approx 2 \pi \times 940\ \textrm{Hz} \approx 4.8 ~
\bar{\omega}$. According to $\bar{\omega}/T_F=1/(3N)^{1/3}$, we get
the ratio between the phonon energy and Fermi temperature
\begin{equation}
\frac{\omega_0}{T_F}=0.045.
\label{eq:omegaoveref}
\end{equation}
From $\omega_0=v_s |\mathbf{q}|$ and $E_F=\frac{1}{2} v_F
  k_F$, we get
$$\frac{|\mathbf{q}|}{k_F}=\frac{1}{2} \frac{v_F}{v_s}\frac{\omega_0}{E_F}.$$
Since in most regime ${v_F}/(2v_s) \sim O(1)$ as shown in
Fig.~\ref{fig:dandv}, using Eq.~\eqref{eq:omegaoveref} we conclude
that  $|\mathbf{q}|$ is also
much smaller than the Fermi momentum $k_F$. Thus, our model, which
requires $q$ to be small, can be applied to this physical system.

\begin{figure}[t]
\includegraphics[width=220pt,height=155pt]{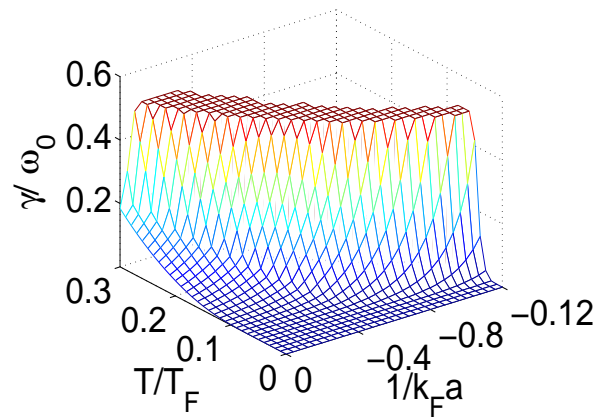}
\caption{(Color online)  Damping rate of collective
excitations in superfluid regime showing dependence on $T/T_F$ and
$1/k_F a$. When $T$ is close to $T_c$,
the results from our approximation are no longer reliable since the
damping rate $\gamma$ is already very large compared with
$\omega_0$. This regime is indicated by the plateau.
Also, when the system is no longer superfluid ($T>T_c$),
our effective field theory does not apply,
and the damping rate is then not plotted there.}
\label{fig:damping}
\end{figure}

Fig.~\ref{fig:damping} shows the numerical results of damping rate
from channel $a$.  It shows that the damping rate
increases for higher temperature and smaller scattering amplitude
$|a|$. In the
superfluid regime, when $T$ is close to $T_c$, the damping rate
$\gamma /\omega_0$ becomes big. When solving the equations for the
damping rate in the preceding section, we assumed a small damping
$\gamma$ compared with the oscillation frequency $\omega_0$,
i.e., kept only the leading order in the perturbative
expansion of $\gamma/\omega_0$.  In the regime of big  $\gamma
/\omega_0$ where a significant correction is expected, our
damping formula is no
longer reliable. Thus, we use a
plateau to indicate the regime of $\gamma / \omega_0>0.5$. Also,
in the normal Fermi Liquid regime $T>T_c$ where our
theory no longer applies, we keep the damping plot open with no data points shown, and will discuss this regime later.

We did not include channel $b$ when plotting
Fig.~\ref{fig:damping}. The reason is that we found that under the
experimental condition \cite{Wright2007a}, the contribution
to damping due to channel $b$ is much smaller than due to channel $a$. To
investigate the features of channel $a$ and channel $b$ in more
details, let us start from the relationship $\omega_q=v_s|\mathbf{q}|$. We
focus in the regime of long wavelength and low temperature such
that $k_F \gg |\mathbf{q}|$, and $T$ is not close to
$T_c$. Thus $\Delta(T)$ is of the same order as $E_F$. In channel $a$,
the first requirement is finite temperature, so that the upper  (lower) band is
populated  by quasi-particles (quasi-holes) due to thermalization. This fact is enforced by the factor
$(n_k-n_{k+q})$ in Eq.~\eqref{eq:imf1tof5a}. The second requirement
is energy conservation. The energy change of the fermionic quasi-particle after scattering with a phonon is as follows,
\begin{equation}
\delta E=E_{k+q}-E_{k} \approx \frac{\xi_k}{E_k} \frac{\mathbf{k \cdot q}}{m_a} \le \frac{\xi_k}{E_k} \frac{|\mathbf{k}| |\mathbf{q}|}{m_a}.
\label{eq:channela}
\end{equation}
$\omega_q \approx \delta E$ is needed to satisfy the energy
conservation. Therefore, we have
\begin{equation}
\omega_q=v_s |\mathbf{q}| \le \frac{\xi_k}{E_k} \frac{|\mathbf{k}| |\mathbf{q}|}{m_a}.
\label{eq:channela1}
\end{equation}

If $|\mathbf{k}|$ is too far away from $k_F$, both the
finite-temperature occupation number and the density of states are
greatly suppressed. Equivalently, the most effective scattering of
phonons is from quasi-particle states around the upper (lower) band's
minimum (maximum). In addition, we have learned in Fig.~\ref{fig:dandv}
that $v_s$ does not exceed the order of $v_F$. As long as
$|\mathbf{k}|$ is of the order $k_F$ or smaller, in principle, phonons
of both large and small $v_s$ may excite fermionic quasi-particles in
channel $a$.  However, the condition Eq.~\eqref{eq:channela1} suggests
that small $v_s$ is much more favored in channel $a$.

In channel $b$, the pair breaking process is allowed by the condition of occupation number at $T=0$ ($n_k=n_{k+q}=0$ in Eq.~\eqref{eq:imf1tof5b}). However, in this case,
\begin{equation}
\delta E=E_{k+q}+E_{k} \ge 2\Delta\,,
\label{eq:channelb}
\end{equation}
where the minimal value of $\delta E$ takes place for $k$ around
$k_F$.
Again, by energy conservation, $\omega_q \approx \delta E$ is
required. Subsequently, we find the minimal condition required of the sound
velocity $v_s$,
\begin{equation}
v_s \ge \frac{\Delta}{E_k} \frac{k_F}{|\mathbf{q}|}v_F.
\label{eq:channelb1}
\end{equation}
As long as $\Delta$ is of the same order of $E_F$, and given that
$k_F/|\mathbf{q}|$ is quite large, the condition
Eq.~\eqref{eq:channelb1} in turns requires that the sound velocity
$v_s$ be much larger than the Fermi velocity $v_F$. In the superfluid
phonon case, the condition $k_F \gg |\mathbf{q}|$ is satisfied, and
when the temperature is still far away from $T_c$, channel $b$ is
prohibited, since $v_s < v_F$ as shown in
Fig.~\ref{fig:dandv}. Another way to understand that the damping
channel $b$ is suppressed is to directly use energy
conservation. Since $\omega_q \ll 2 \Delta$, channel $b$ can not
happen.

At finite temperature, channel $a$ becomes possible since some
fermionic quasi-particles and holes are thermally created in the upper
and lower band, respectively. They can scatter with superfluid phonons
to cause decay. However, channel $b$ is still greatly suppressed as
long as $w_q < 2\Delta$.

In the experiment, Since $\omega_0/E_F \approx 0.04$, channel $b$ can
happen only when when $\Delta(T)/E_F \approx 0.02$. From mean field
analysis, near $1/k_F a \approx -0.5$, to satisfy $\Delta(T)/E_F
\approx 0.02$, we need to have $T/T_c \approx 99.7\%$, which is very
near the phase transition. However, according to
Ref.~\cite{Wright2007a}, the damping happens at $T \approx 0.6 T_c$
near $1/k_F a \approx -0.5$, which is too low to let channel $b$
happen from our calculation. Even if one uses some theories including
quantum fluctuation \cite{PhysRevA.75.033609}, at $T \approx 0.6 T_c$,
$\Delta(T) / E_F$ is still much larger than $0.02$. Thus the damping
peak should not correspond to the pair breaking channel $b$. As $T$
getting closer to $T_c$, channel $a$ will be more and more enhanced
because of more and more thermally excited fermionic
quasi-particles. When $\Delta(T)/E_F \approx 0.02$, channel $b$ also
happens, while the damping from channel $a$ is already
very large. It is not clear which channel dominates because $T$ near
$T_c$ is outside the valid regime of our low energy effective field
theory. It remains to be a challenge to formulate a quantum theory
beyond the classical Boltzmann equation. Thus, our calculation just
considered the contribution from channel $a$. We also double checked
channel $b$ by numerical method and confirmed that the contribution
from channel $b$ is zero for most regime.

There is another experimental evidence showing that why channel $b$
does not dominate. In the experiments varying the magnetic field
\cite{Bartenstein2004, Altmeyer2007}, if the pair breaking mechanism
had dominated, one should also have observed very sharp peaks in these
experiments. However, the damping rate changes relatively smoothly
\cite{Bartenstein2004, Altmeyer2007}, which means channel $a$ should
be the reason for the smoothly increasing damping. Therefore, we
expect the damping observed by changing the temperature
\cite{Wright2007a} to be due to channel $a$ too, since increasing
the magnetic field at a finite temperature has the same effect as
increasing the temperature at fixing magnetic field, both just
reducing the gap. Nevertheless, the pair breaking channel $b$ is also
possible to happen near the phase transition, but it is not necessarily dominating in contrary
to what has been suggested.

The mechanism discussed here is different from the conventional
acoustic attenuation process in solid-state
superconductors, where channel $b$ overwhelms channel $a$ and
$\omega_q \approx 2\Delta$. For the conventional case, $v_s$ is big,
so that a relatively large phonon energy $\omega_q$ corresponds to a
very small momentum $\mathbf{q}$. Channel $a$ is suppressed as $v_s$
is too big to satisfy the energy conservation condition
\eqref{eq:channela1}. At the same time, a channel $b$ pair breaking
process (Fig.~\ref{fig:channel}(b)), associated with small momentum but
large energy transfer from acoustic phonons to Fermi quasi-particles,
may happen. The original fermion pair breaking process is the creation
of a pair of particle and hole in the upper and lower branches of
quasi-particle energy spectrum, respectively. A small $|\mathbf{q}|$
ensures that fermionic quasi-particles are created at the band
extrema, which is known to result in a peak of damping rate,
$\gamma\propto 1/\sqrt{\omega-2\Delta}$, due to the singularity of
density of states. Therefore, the damping for large $v_s$ in
traditional case is due to the pair breaking process (channel $b$).

We did not consider the fact that in the experiment, the fermionic
gases are trapped and inhomogeneous in space. The above calculation is
effective for gas at the center of the trap. On the edge of the
trapped gas, by local density approximation, the effective density and
Fermi energy is smaller than that of center, which means $\omega_0 /
E_F$ is larger and channel $b$ may happen in lower temperature (but
still not very low yet). However, channel $a$ also happens on the edge
since the above analysis still works for lower gas density on the edge,
and the effect of channel $b$ is just to increase the damping rate,
not giving a sudden peak.

\begin{figure}[t]
\includegraphics[width=230pt,height=200pt]{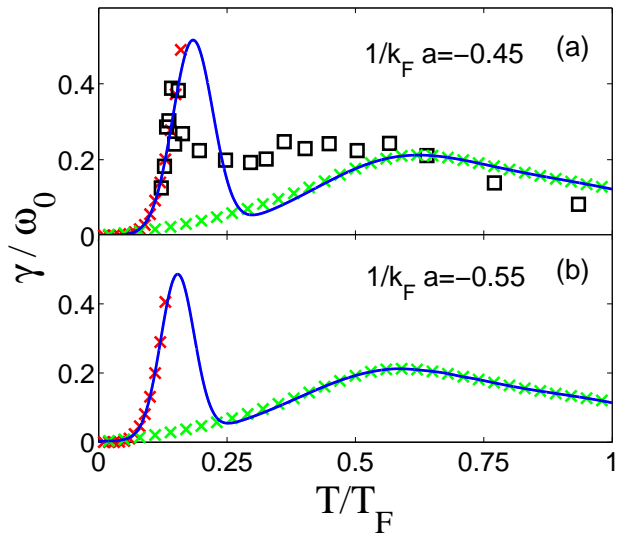}
\caption{(Color online) Damping rate of collective excitations by interpolating results from our effective field theory in low temperature and the Boltzmann equation in high temperature. Red crosses are the prediction from our calculation in low temperature superfluid regime, and green crosses are from the Boltzmann equation. The blue solid line is the interpolation. The black squares are the experimental data of damping~\cite{rgdata}. 
(a) $1/k_F a=-0.45$; (b) $1/k_F a=-0.55$. The first peak moves toward higher
temperature when the system gets closer to the resonance (i.e., smaller $|1/k_F a|$).}
\label{fig:twop}
\end{figure}

In the regime $T>T_c$, where the system is no longer superfluid and
our effective theory breaks down, the classical Boltzmann equation can
be used to calculate the oscillation frequency and damping rate, as
shown in Ref.~\cite{PhysRevA.76.045602}. As supplement to our main
result above, we use the equation in Ref.~\cite{PhysRevA.76.045602} to
calculate the oscillation frequency. As mentioned there, the formulas
did not take into account the effect of Pauli blocking, which means
that the classical Boltzmann results are more valid at
high temperatures. The
calculation takes into account the trapping potential. At $1/k_F
a=-0.45$, if we interpolate our theory in the low temperature
superfluid regime with the results from Boltzmann equation in the high
temperature Fermi liquid regime, two peaks by two different methods
appear in Fig.~\ref{fig:twop}, which agrees to the experimental
results fairly well (cf. Fig.~2(b) in Ref.~\cite{Wright2007a}). Thus, we
conclude that the first sharp peak observed in Ref.~\cite{Wright2007a}
is due to the superfluid phonon and fermionic quasi-particle
interaction,  mostly through channel $a$ in
Fig.~\ref{fig:channel}(a). The second broad peak is given by the
Boltzmann equation
from Ref.~\cite{PhysRevA.76.045602}, which signals the hydrodynamic
to collisionless transition. Our model further shows that the first damping peak moves toward higher
temperature when the system gets closer to resonance (i.e., smaller $|1/k_F a|$), as one can see in the change from Fig.~\ref{fig:twop}(b) to Fig.~\ref{fig:twop}(a). This phenomenon was first reported in the experiments of Ref.~\cite{Wright2007a} (cf. Fig.~3 therein). In summary, our theory provides a consistent explanation for the experiments of damping.

According to our calculation, a damping due to phonon-fermion
interaction should happen in the unitary limit. While this was not
reported in the experiments of
Ref.~\cite{Wright2007a}, in the experiments of Ref.~\cite{PhysRevLett.94.170404}, the authors found that the damping rate of a Fermi gas at unitarity displays a weak peak immediately followed by a notch near transition as temperature increases.  Such a damping notch is consistent with the dip of Fig.~\ref{fig:twop} that we propose here.

There are several reasons for the quantitative discrepancy between our
calculation and the experiments. The most important thing is that our
calculation is based on mean field results of $\Delta_0$ and $\mu$
from solving Eq.~\eqref{eq:mfreg2}. Quantum fluctuations intend to
destroy the superfluid phase, i.e., reduce the transition temperature
$T_c$ below the mean field results. More reliable inputs for
$\Delta_0$ and $\mu$ from calculation including quantum fluctuations
\cite{PhysRevA.77.023626} will give a lower $T_c$, which will make our
results better agree with experiments. Secondly, our calculation is
based on Fermi gas in free space while in the experiments, the gas is
always trapped.  Thirdly, we used approximation to solve the
damping. However, to be strict one needs to exactly solve for the
poles. Also, $\omega_0$ is read from Ref.~\cite{Wright2007a} and
treated as a constant in our work. That is just an approximation since
$\omega_0$ is also changing slightly with temperature.

To verify our main conclusion that channel $a$ is the dominating process, we propose an experiment, which is to measure the damping rate $\gamma$ while varying the oscillation frequency $\omega_0$, and observe how $\gamma / \omega_0$ changes. This can be done by either increasing the particle number or reducing the trapping potential. If channel $b$, the pair breaking mechanism, is dominating, one should observe that the damping peak becomes increasingly narrower, and eventually is too narrow to be observable, as $\omega_0$ decreases. The reason is that the parameter regime to satisfy the energy conservation $\omega_0\approx2\Delta$ diminishes with decreasing $\omega_0$, and eventually vanishes.  On the other hand, if channel $a$, the Landau damping mechanism, is dominating, no matter how small $\omega_0$ is, as long as Eq.~\eqref{eq:analytical1} is satisfied, the energy conservation law is always satisfied. Eventually $\gamma/\omega_0$ will be independent of $\omega_0$ in the limit of vanishing $\omega_0$ as discussed before in Eq.~\eqref{eq:gammaprop}, and therefore the peak of damping $\gamma / \omega_0$ remains unchanged and should always be observable.

To summarize, we have studied the damping of collective modes for ultracold Fermi gases by effective field theory, and compared our results with experiments. By varying the temperature \cite{Wright2007a} our theory predicts that one should find two peaks of damping. The first is a sharp peak due to the finite-temperature phonon-fermionic quasi-particle interaction, and the system changes from hydrodynamic superfluid to hydrodynamic Fermi liquid. The second is a broad peak due to the transition from the (collisional) hydrodynamic to collisionless Fermi liquid (see Fig.~4 in  Ref.~\cite{Wright2007a}). In the experiment that varies the magnetic field \cite{Bartenstein2004, Altmeyer2007} but keeps temperature sufficiently low, one should see only the peak due to the phonon-fermionic quasi-particle interaction. After passing
this peak, the system is already in the collisionless regime without the
need of going across a collisionally hydrodynamic regime. Thus, there
is no second broad peak.

\section*{ACKNOWLEDGEMENTS}
We thank R. Grimm for sending us the experimental data. We thank M. Urban for comments and suggestions. We also thank J. Thomas and H. Heiselberg
 for helpful discussions. This work is supported in part by U.S. Army Research Office Grant
No. W911NF-07-1-0293 and the CAS/SAFEA International Partnership
Program for Creative Research Teams of China.

\bibliography{Damping}
\end{document}